\documentstyle[12pt]{article}
\begin{document}
\begin{titlepage}
\title{
The Concept of Mass, Quarks, and Phase Space}
\author{
{P. \.Zenczykowski $^*$ }\\
{\em Dept. of Theoretical Physics},
{\em Institute of Nuclear Physics}\\
{\em Polish Academy of Sciences}\\
{\em Radzikowskiego 152,
31-342 Krak\'ow, Poland}\\
e-mail: zenczyko@iblis.ifj.edu.pl
}
\maketitle
\begin{abstract}

We point out the conceptual problems related to the application of
the standard notion of mass to quarks  and recall 
the arguments that there should be
a close connection between the
properties of elementary particles 
and the arena used for
the description of classical macroscopic processes.

Motivated by the above and the wish to introduce more symmetry between the
coordinates of
position and momentum we concentrate on
the classical nonrelativistic phase space with ${\bf {p}}^2+{\bf {x}}^2$ 
as an invariant. 
A symmetry-based argument
on how to generalize the way in which mass enters into
our description of Nature is presented 
and placed into the context of 
the phase-space scheme discussed.

It is conjectured that the proposed non-standard way of
relating "mass" to the variables of the classical phase space
is actually used in Nature, and that it manifests itself through
 the existence of quarks.
Some properties of this proposal, including unobservability of "free quarks" 
and the emergence of
mesons, are discussed.

\end{abstract}

\end{titlepage}

\section{Introduction}

The present paper has its roots in a long-held 
suspicion that the discrete quantum properties of quarks and other elementary
particles 
are closely linked to the properties of a continuous arena 
used for the description of classical macroscopic processes.
This arena is usually thought to be provided by the spacetime.
However, as in physics one deals with Nature only through a language 
chosen for its description,
this arena may
depend on the language chosen.  
Thus,
as the present paper is concerned with the idea of introducing more symmetry
between the position and momentum coordinates of classical nonrelativistic
physics, the arena in question appears to be that of classical phase space, and
the language that of a Hamiltonian formalism, in which momentum and position are
treated as independent variables.

Some aspects of this idea were considered already in 
ref. \cite{ZenczyIJTP1990}, where it was pointed out that
 as the group of rotations in the three-dimensional space,
 treated as an automorphism group of an underlying algebra, singles out the
 algebra of quaternions, 
 so the group of transformations in the six-dimensional phase-space under some
 additional assumptions
 points to the algebra of octonions.
 With the physical meaning of octonion nonassociativity fairly obscure,
 this route does not lead far at present, however.
 
 In
the following the idea of introducing more symmetry between position and
momentum is pursued at a familiar associative level, tailored to admit
a fairly clear interpretation of at least some points.

The paper is organized as follows. In the next Section various problems with 
the standard concept of mass as applied to quarks are presented.
In this way the observational basis suggesting the need for a different
approach to the issue of quark mass is laid out.
Section 3 provides general arguments for a close connection between the
properties of elementary particles and the classical arena
on which these actors are thought to play. 
Then, phase-space symmetries are considered in some detail, 
and an argument is presented on how to generalize the way in which
mass enters into our description of Nature. 
It is then conjectured that the proposed non-standard way of
relating the concept of mass to macroscopic variables
is actually used in Nature, and that it manifests itself through
 the existence of quarks.  
 In Section 4 the arguments of the preceding Section
 are used to generalize the Hamiltonian of a standard spin-1/2 particle
to the case of a non-standard mass concept.
 Then, charge-conjugation is discussed 
and a simple-minded treatment of a quark-antiquark system is presented.
Section 5 points to a 25-years-old scheme with composite leptons and quarks,
 which exhibits vague (even though superficial) similarity to our proposal. 
 Our final remarks are contained therein as well.

\section{The Problem of Mass}
 
	The standard concept of mass, 
originally introduced for the description of large classical objects 
moving along well-defined trajectories and responding in a well-defined way 
when subject to external forces, 
	may be applied to free individual elementary particles 
such as leptons, photons, or hadrons
when appropriate care is taken to account for their quantum properties.
				
	Validity of this extrapolation 
from large and classical to small and quantum
is in particular confirmed by the successes of quantum electrodynamics (QED),
our best microscopic theory, 
when applied to	electron, muon or the $\tau$ lepton. 
				
	However, while this theory permits us to predict 
some experimental numbers (such as e.g. lepton magnetic moments) 
with astonishing accuracy, thus corroborating in detail
the structure of the theory, 
and in particular the way mass enters into it, it does not say anything about 
the ratios of lepton masses.
In fact, the question "Who ordered that?" asked by I. Rabi 
when muon was discovered over half a century ago, 
remains completely unanswered in the 
nowadays widely accepted Standard Model (SM) of elementary particles, in which		
	the appearance of several types of leptons and 
the problem of their masses constitute a part of a larger puzzle.

\subsection{Free Parameters of the Standard Model} 

The Standard Model has had many spectacular successes 
in correlating and explaining a vast body of elementary particle data.
However, 
being in its very essence a
non-Abelian generalization of the Abelian theory of quantum electrodynamics, the
Standard Model 
inherits the limitations of the latter, thus
in particular being unable to provide any prediction 
for the ratios of lepton masses.
In addition to lepton masses,
it also contains several further parameters
with values completely unexplained:
	e.g. quark masses, mixing angles and phases.
Furthermore, in the realm of strong interactions,   
the accuracy of the agreement of SM predictions (or SM-inspired expectations)
with experiment is in general significantly lower.

	These shortcomings of the Standard Model 
	may be viewed as indicating 
	that the solution to the problem of 
why elementary particles have these or other mass values etc. may require New
Physics. At present this term is used mainly to denote any kind of
departure from the Standard Model, tacitly assumed to be describable 
within a field-theoretic approach.
However, with no such departure having been identified as yet,
nothing is really known about "New Physics".
Thus, the latter
 may also involve going beyond the realm of quantum field theory. 	
				 
	Now, the SM quark mass parameters are introduced 
into the fundamental Lagrangian in a complete analogy with lepton masses.
In its essence, therefore, the SM quark mass is a standard classical concept.
Other SM parameters (e.g. mixing angles, phases) 
are more quantum-like in their
nature.
				
	The present paper deals with the conjecture that the extrapolation of
the standard classical concept of mass to quarks, assumed in the SM as an input, 
may not be wholy warranted. 
	As pointed out above, the concept of mass was originally introduced 
for the description of the behaviour of {\it individual} classical particles; 
with appropriate shift to the language of quantum physics 
the use of this concept
 was then extended to the description of the behaviour of 
{\it individual} elementary particles 
observed {\it far away} from the region of interaction.
	Quarks, however, are supposed to be permanently confined in hadrons
	and have never been observed as individual particles 
	far away from the region of interaction.
	Thus, the applicability of 
	the standard concept of mass to quarks may be questioned.
	I believe that the fact that free quarks have not been observed 
hints that quark theories should not 
be based on the standard concept of mass.  
This belief is supported by conceptual problems encountered in the Standard
Model when the standard concept of mass is applied to quarks. 
These problems shall be presented in Section \ref{SMproblems}.

	In order to avoid
	imminent charges that the advocated point of view denies
the applicability of the concept of mass to quarks,
it should be strongly stressed right from the beginning 
that it is accepted herein 
{that a parameter of the dimension of mass 
and therefore {\em some}
concept of mass
may be assigned to a quark. 
	Indeed, despite the fact that separate individual quarks 
	have not been observed
(and consequently that their mass cannot be directly measured 
in an experiment similar to one that can be used for leptons or hadrons),
	there are many indirect indications in favour of this view.
	What in my view remains open is the question whether
the direct, straightforward application to quarks 
of the {\it standard} classical concept of mass, 
together with all implicit
properties of   such a mass, 
constitutes the proper extension of the latter concept 
into the subhadronic world.
	I think that the appropriate question here is: 
	"what is the proper way of assigning a 
parameter of the dimension of mass to a quark?". 
In other words, I think that the present way of
extrapolating the standard concept of mass to quarks
requires modifications.

	In fact, for many years there was a heated 
	debate whether quarks should be conceived 
	of as ordinary particles or as mathematical 
	entities permitting a successful description
	of hadrons. 
	    Development over the years proved 
	that quarks interact in a point-like manner, much like the leptons, 
	thus shifting the concept of a "quark" more and more 
	into the realm of "ordinary" particles.
	However, there still remains the question 
how much of this shift is really necessitated by
the experimental information available, 
and how much appears to represent our unjustified expectation only.
	Interaction of quarks in a point-like manner, 
	 and 
the applicability to quarks of the classical concept of mass 
constitute two separate issues. 
Therefore, the latter should not be assumed on the basis of the former. 		   
    Accordingly, 
    following the general view that the issue of masses constitutes a weak point
    of the SM (or rather a point beyond that scheme), and
    in line with the arguments presented above, 
    I believe that the SM requires
    some reinterpretation as far as quark masses are concerned. 
    				    
	It is hoped that the argument later on, 
and in fact the whole paper, will clarify my views in the matter, provoke
some feedback, and perhaps help in the development of
 the ideas presented below.

\subsection{Masses from Higgs Mechanism}
In the Standard Model, all elementary particles
acquire their masses through the Higgs mechanism. 
One cannot consider
	this mechanism as representing a solution
	of the problem of  mass as defined above 
	(i.e. explanation of the ratios of elementary particle masses). 
	Indeed, the Higgs mechanism merely shifts the problem to that of 
	the interaction of Higgs particles with matter fields. 
	There is no apparent bonus resulting from this shift: 
	the couplings of Higgs particles to matter fields 
	remain as unconstrained as the masses and other parameters
 	of the Standard Model.
 
Although the Higgs mechanism constitutes 
a crucial ingredient of the Standard Model,
allowing for renormalizability with massive gauge bosons, 
it has its shortcomings.
On the theoretical side, it generates a constant term which contributes to the
energy density of the vacuum 55 orders of magnitude larger than observed.
On the experimental side, Higgs particle has not been observed as yet.
On the philosophical side, introduction of such a particle may be viewed as
resulting from human mind's tendency to explain the natural
phenomena in terms of material
objects rather than in terms of abstract concepts, a tendency that has misled us
several times in the past. 
Thus, the introduction of Higgs particle(s), 
	while solving a problem in the description of weak interactions,
	is regarded by many physicists with some suspicion.

In fact, attempts to dispose of Higgs particle became more often recently.
There are proposals in which no such particle is present \cite{noHiggs}.
If Higgs particle is not found soon, 
alternative approaches will certainly be pursued more often.
Conversely, even if Higgs particle were found, 
we would be still very far from the solution of the problem of mass.  

New hints on how to deal with the issue of mass could perhaps be unveiled
through a search for
 regularities in mass matrices for leptons and quarks.
Interesting attempts to derive 
lepton and/or quark mass matrices from some underlying Ans\"atze
were made \cite{massmatrices}. 
Some of these papers accept implicitly 
that the introduction of quark mass "\'a la lepton mass" is fully legitimate.
However, as already hinted at above,
I think that it brings about serious conceptual problems,
to be exposed in more detail below.
  
\subsection{Problems with Quark Mass}
\label{SMproblems}

	Let us return to the concept of quark mass and how it enters 
	the present-day theories. 
	Within the Standard Model quark interactions are described by
	quantum chromodynamics (QCD), 
	the non-Abelian gauge theory of interacting quarks and gluons. 
	The concept of quark mass enters here at the level of
	Lagrangian in much the same way as in quantum electrodynamics.
	Given the enormous success of quantum electrodynamics, 
	the mathematical elegance of QCD - a non-Abelian generalization of QED,  
	and the asymptotic freedom property of QCD,
	so appropriate for the description
	  of characteristic features
	observed in the scattering at large momentum transfers 
	in terms of hadron constituents,
	this way of introducing quark masses seems completely natural.
	
		A problem appears, however,
		when one attempts to make a connection 
		between quark masses introduced in the way just described, 
		and quark masses extracted in various ways 
		from the experimental data.
		Since individual quarks are not observed,
		any such extraction must involve a great deal of theory.
	
	However, there is no good theoretical way to make the connection
	in question.
	Quark masses appearing in the Lagrangian 
	(the so-called "current" quark masses, especially those of the light
	quarks)
	cannot be extracted from data at large momentum transfers, where
	theory is under better control, for two reasons. First, 
	in this region these masses are negligibly small in comparison to the 
	momenta in
	question. Second, we do not know how to make a theoretically sound
	(i.e. based on the Lagrangian only)
	connection between quarks
	and the experimentally observed hadrons.
	Because of the first obstacle 
	quark masses have to be (and are) extracted from low-energy and 
	low-momentum transfer data.  
	However, it is widely agreed that in this region the second obstacle
	is particularly troublesome and that
	the perturbative approach to strong interactions of quarks 
	must become invalid on account of confinement effects, 
	with nonperturbative effects becoming dominant. 
	Unfortunately, since the latter effects are not calculable 
	they could affect the extraction in an unknown way.
	In order to proceed,
	one has to make here additional assumptions of phenomenological nature, 
	thus leaving the solid ground of fundamental theory, and admitting other
	ingredients besides the First Principles.
	At the very least, therefore, 
	the values of quark mass parameters extracted in this way are 
	uncertain in a serious  way. 
	This prevents sound confirmation of the concepts used
	and opens doubts as to the meaning of the parameters extracted.
	
	The simplest way of proceeding (and the one used very often) 
	is to essentially forget about the confinement of quarks 
	and to perform the calculations perturbatively 
	using Dirac bispinors for external quarks 
	and standard Dirac propagators for internal quarks.
	This is being done in the quark parton model 
	and in all perturbative Standard Model calculations, 
	bringing in a serious conceptual problem.
	Indeed, the pole present in the standard Dirac propagator, 
	occurring at momentum square equal to quark mass squared 
	and corresponding to a free classical particle, 
	should be {\it absent} in the calculations
	as free quarks (in asymptotic states)
	are not observed in the real world.
	Consequently, one wonders what is the meaning of all perturbative 
	calculations in which standard Dirac propagators and bispinors
	are used for quarks.
	Yet, it is precisely through such calculations 
	with standard Dirac quarks and their propagators 
	that information about quark masses is being extracted from the data.
	Below we briefly recall a few important calculations of this type.

\subsubsection{{External quarks (in initial/final hadrons)}}

{\it Current quark masses} \\
	These are the masses used in the fundamental Lagrangian.
	For example, the ratios of the current masses of light ({$u,d,s$}) 
	quarks 
	were extracted from the combinations of the 
	masses of kaons and pions  
	\cite{currentmasses}
	using standard Dirac bispinors for external quarks. 
	The relevant procedure considers quark currents, and  
	in particular the axial current $\bar{q} \gamma _{\mu} \gamma _5 q$, 
        applying to its divergence
	the Dirac equation $\hat{p}_{\mu}\gamma ^{\mu}q=m_{current}q$. 
	This leads to expressions
	proportional to the sums of current quark masses, and the possibility of
	extracting them from the data under some additional assumptions. 
	Note, however, that using the Dirac equation means that we adjust quark
	momenta and put	quarks on their mass-shells, as appropriate for free
	particles. 
	 With momentum well-defined only when the associated
	wave is sufficiently long in ordinary space,
	one should expect this procedure to be
	invalid for quarks "confined to a small region of space within a
	hadron". 
	Yet, it is such treatment of currents which, 
	when further ingredients are added, leads from hadron masses 
	to the nowadays widely accepted low values of current masses 
	for the $u,d,s$ quarks,
	with $u,d$ current masses
	of the order of a few $MeV$.
	While there is no doubt that important information is being extracted
	in this way from the data, its meaning is blurred by
	the conceptual inconsistency indicated above,
	raising  in particular the question of the
	meaning of the mass(-like) parameters thus extracted.\\

	{\it Constituent quark masses}\\ 
	In many papers the calculations were performed using the so-called 
	"constituent" quark masses. These are not the masses present in the
	underlying Lagrangian, but the effective masses approximately equal to 
	one third of the mass of an appropriate baryon, 
	or a half of the mass of the corresponding vector meson.
	They are thought to take into account in some effective way
	the effects of the confining interactions.
	Thus, the effect of the confining interaction is thought to contribute
	some $330-350~ MeV$ to the effective quark masses.
	While these masses, being defined independently from the Lagrangian,
	do not have such a strong theoretical basis
	as the current masses, they work equally well
	in their respective domain of applicability.
	
	Originally, the constituent quark masses appeared 
        when the magnetic moments of octet
	baryons (i.e. proton, neutron etc.) were considered.
	Amazingly, the assumption of constituent quarks described by Dirac
	bispinors and treated as {\it free} particles (i.e. with 
	$p_{\mu}\gamma ^{\mu}q=m_{const}q$), when
	supplied with the appropriate symmetry structure of 
	three-quark baryonic states 
	and the principle of additivity of contributions from
	the individual quarks resulted in a very good description of baryon
	magnetic moments.
	One faces here a conceptual problem again: the treatment of constituent 
	quarks as free Dirac particles 
	leads to good results despite the fact that individual
	free quarks are not observed, and that constituent quarks are
	supposed to describe confined quarks.
	Many attempts were proposed in order to overcome such
	objections. 
	For example, within the approach of a bag model, the size of
	magnetic moments is set by the size of a bag to the
	interior of which the quarks are confined. Despite such attempts,
	no improvement in the
	description of magnetic moments followed.
	 In fact, the best parameter-free model for baryon
	magnetic moments is that of Schwinger \cite{Schwinger}, 
	in which the size in
	question
	is set by the (experimentally measured)
	masses of vector mesons, with the rest 
	of the success ensured by the spin-flavour 
	symmetry of baryon wave functions alone.

\subsubsection{{Internal quarks}}

Perturbative treatment of the Standard Model leads to the appearance of quark
propagators which may occur in loop and/or tree
diagrams. The resulting formulae depend on quark masses
present in these propagators.\\

 {\it Masses in loop propagators}

	An example of  a calculation depending on the mass
	present in loop propagators only is furnished by
	the prediction of the charmed quark mass.
	This mass was predicted
	through the use of box diagrams with internal charmed quarks
	 \cite{MKGaillard}, and applied to uncharmed external states.
	When experimental data were combined with the formulae
	obtained in this way, a mass value of about 1.5 - 2.0 $GeV$ was 
	predicted. This mass was believed to be mainly the current mass, 
	as only small ($330-350~MeV$ or 20\%) corrections from 
	confining interactions were expected.  
	Subsequent experimental discoveries of charmed particles
	with effective charmed quark mass around 1.5 GeV provided arguments in
	favour of the applicability of such calculations.

	Since the current masses of the $u,d,s$ quarks, determined as described
	earlier, are small 
	when compared to the $330-350~MeV$ scale, 
	for these quarks it is the constituent quark masses that are
	expected to be appropriate for loop calculations. In fact, the
	constituent masses of "light" quarks were successfully 
	used in a series of loop 
	calculations of meson formfactors 
	 \cite{Bramon}. \\
	
 {\it Masses in tree propagators}

	The results of loop diagram calculations suggest that quark
	poles are present. 
	The unobservability of quarks 
	suggests, however, that these poles should be absent.
	This contradiction can be studied in tree diagrams, where the 
	existence of quark poles 
	may be directly tested through the brehmstrahlung process. 

		The question if it is justified to use 
		standard Dirac
		propagators for intermediate constituent quarks in photon
		brehmstrahlung 
		was resolved as a by-product of the
		studies of weak radiative decays of hyperons (WRHD). 
		
		Calculations of the WRHD amplitudes 
		in the framework of the constituent quark model
		(only $u$, $d$, or $s$ quarks were involved)
		appropriately generalized to describe 
		weak interactions,
		with poles due to intermediate constituent quarks
		and thus testing the meaning of the
		constituent-quark description of baryon
		magnetic moments somewhat deeper,  
		have resulted in the firm prediction 
		of a nearly maximal positive asymmetry in
		the $\Xi ^0 \to \Lambda \gamma $ decay \cite{LachZen}.
		This turned out to be 
		in complete disagreement with a
		nearly maximal negative asymmetry found later in experiment
		\cite{NA48WRHD}.
		
		The origin of the incorrect prediction has been traced to the
		presence of the constituent quark pole. 
		A correct description \cite{ZenWRHDresolution} 
		of experimental data on WRHD 
		requires abandoning the treatment of constituent quarks \'a la
		leptons, i.e. with mass-involving standard propagators,
		in favour of Gell-Mann's concept of $SU(3)_L \times SU(3)_R$ 
		symmetry of currents \cite{GellMann}, involving
		in its bare form no quark masses at all, 
		in clear analogy to the Schwinger's
		 treatment of baryon magnetic moments \cite{Schwinger}.
		
		In effect, weak radiative hyperon decays {showed} that the
		treatment of constituent quarks 
		as ordinary Dirac particles with standard propagators, 
		is generally 
		incorrect as it leads to 
		a sharp disagreement with experiment, i.e. to artefacts. 
		What remains of the
		constituent quark approach is the description of external baryons  	
		in terms of an appropriate spin-flavour wave function, with the
		dominant role played by its symmetry, and with the concept of
		constituent quark mass understood (or better: defined) just 
		as a half of the corresponding
		vector meson mass, in agreement with the original ideas of
		Schwinger \cite{Schwinger}.}\\

		The situation presented above is deeply dissatisfying, 
		not only because
		of the existence of
		the conceptual problems and internal inconsistencies 
		related to the use of quark mass, but also because usually
		they are either unnoticed or swept away
		"under the carpet". 
		
		There are some hints, however.
		Namely,
		it seems that 
		whereas the use of external quarks endowed with standard
	        mass leads to conceptual problems 
		and, similarly, the use of standard quark propagators in
		tree diagrams results in artefacts,
		in 
		loop calculations the
		concept of poles associated with quarks seems to work, 
		at least effectively and to some extent.
		
		This suggests that the concept of quark mass 
		(and/or the related pole) 
		should undergo some reinterpretation 
		so as to still allow for some poles (or their analogues) 
		in the loop diagrams 
		while clearly forbidding standard quark pole contributions 
		in the tree diagrams. 
		This could be achieved e.g. 
		by admitting the concept of mass to
		hadrons only (consider e.g. \cite{Steinberger}, or the idea of 
		hadronic bootstrap supplemented with some quark degrees of
		freedom).
		On the other hand,
		if mass parameters are to be assigned more directly
		to quarks,
		it does not seem appropriate
		to lay fault with the dynamical confinement of quarks alone 
		as the latter should
		kill the poles everywhere, 
		i.e. both in the tree and in the loop
		amplitudes.
		In the following we will present an idea, induced by
		considerations outside of the Standard Model, which 
		forbids the appearance of standard poles by construction while
		still admitting for quarks the concept of a mass parameter
		and which, as we hope,
		could shed a different light on the issue of
		quark mass.
		
\section{Mass and Space}
\label{generalization}
\subsection{Fundamental Mass / Parameters}

Ultimately, the resolution of the problem of mass requires finding a theory 
which would assign dimensionless numbers to all mass ratios.
In order to predict particle masses,
  this theory should be
accompanied by at least one additional constant 
of the dimension of mass, which would
 set up the mass scale. 
(The Higgs appproach fits into this philosophy as well, one of 
its problems lying in the said prescription for dimensionless numbers).

This constant then plays the r\^{o}le of a "fundamental" mass.
Alternatively, the latter may be constructed from other parameters of the
theory, deemed to be "more fundamental". 
Then, the relevant parameters do not have
to have the dimension of mass themselves.

\subsection{Quantum Numbers and Space}

The origin of the proliferation of elementary particles 
and their quantum attributes
constitutes a problem untouched by the Standard Model. 
The variety of particles and their quantum numbers, 
mass 
parameters, etc., are
duly taken into account in the SM,
but remain completely unexplained 
in terms of any simple single underlying principle.

	When searching for such a principle one should note that
	some quantum attributes of elementary particles 
	are clearly associated with the properties of 
	classical macroscopic continuous space
	in which these particles are envisaged as "moving". 
	One can single out here the concept of particle spin 
	and the corresponding classical notion of macroscopic rotation.
	Likewise, the notion of chirality is related to the existence of 
	left-right symmetry in the macro-world.
	Similar connection between properties of particles and space 
	exists also in general relativity, 
	where matter (i.e. particles) defines and modifies
	the properties of space.
	
The above examples suggest that other particle attributes 
(e.g. particle types and their quantized masses)
should also be somehow
connected to the properties of the macroscopic space.
In fact, it is philosophically very tempting to
conjecture that this should hold for all particle attributes.
According to this philosophy,
properties of particles should correspond to properties of space
and vice versa. 
	Many attempts were pursued along that way.
	Attempts to unite Poincar\'e 
	and internal symmetries were shown to lead to a "no-go" situation. 
	However, what any such no-go theorem really proves 
	is  that the one particular way (or class of approaches) 
	considered in the theorem is not acceptable. 
	All other possibilities of connecting space and particle properties 
	(conceptually perhaps completely different) 
	are not restricted in any way 
	and remain fully open until proven otherwise.
	
Arguments have also been presented that
the macroscopic continuous classical space and the elementary particles
themselves should not only be closely connected with each other 
but actually constructed from a deeper level - 
the underlying quantum pregeometry
 \cite{Penrose68,Penrose71,Wheeler1973}.
While not pursuing such arguments in detail,
I accept their resulting philosophy, i.e. the conclusion 
that space and particle attributes should be intimately related,
and
that to any given particle attribute there should correspond a
certain attribute of classical macroscopic arena
used as a background for the description of physical processes.
In this context it is appropriate to quote
here the words of Penrose:
"I do not believe that a real understanding of the nature of elementary
particles can ever be achieved without a simultaneous deeper understanding of 
the nature of spacetime itself" \cite{Penrose68}.

Now,  
it was argued above that
theory should forbid the appearance of standard
quark propagators in the tree diagrams,
while still admitting
some meaning to their presence in the loops.
The simplest way to satisfy the first condition is to forbid standard quark
propagators 
 right from the beginning.
 If one still wants to assign some concept of mass to quarks, 
 one has to detach the concept of quark mass from that of the standard pole.
 This seems to require a broader theoretical structure.
Consequently,
I think that this provides us with a hint
that one needs to use a broader
 concept of underlying macroscopic space.

\subsection{Mass and Phase Space}

 Since our intuition seems to work best 
 at the classical macroscopic level, it is at this level, as I believe,
 that
 the generalization procedure should be started, with 
the world of quantum attributes to be reached only later.

As noted in the Introduction, the arena used
for the
description of classical processes depends on the language chosen.
If one chooses here the Hamiltonian formalism,
it is the phase space - with independent position and momentum coordinates -
which becomes the arena.
The change of background from ordinary space to phase space seems to be 
the simplest generalization
possible.
 
Now, let us note that the description 
of Nature in terms of the concept of spacetime 
as the arena on which processes may be visualised,
required the connection 
of such manifestly different notions as
space and time into a single construct.
In the end this was achieved in particular 
through the use of the velocity of light $c$ as a
dimensional constant permitting time and space to be measured 
in the same units so that they may be transformed into each other,
as suggested by the properties of Maxwell equations.

Keeping in mind the conjectured connection
between the properties of elementary particles and the classical
 background needed
for their description, the
string-like properties of hadrons suggest
 an introduction of a different 
dimensional constant $\kappa$
of dimension [$GeV/cm$],  which would make it
possible to treat momentum and position in a more symmetric manner, and admit
their mutual transformations. 

In a future quantum theory, yet to be constructed,
the presence of both this constant and the Planck constant
 should be sufficient to set the scale 
for all quantized masses.
For the time being, however, I will consider momentum and position as
commuting variables. 	
		Over 50 years ago 
		the idea of introducing more symmetry
		between the canonically conjugated position and momentum 
		coordinates of nonrelativistic physics
		 led M. Born \cite{Born1949}
		to his reciprocity theory of elementary particles, in which he
		considered the concepts of fundamental length and fundamental
		momentum, and introduced the
		 "reciprocity" transformations 
		$x \to p$, $p \to  -x $.
		
		The first decision one has to made 
		when introducing $\kappa $ is to decide 
		whether one should begin with 
		the four-dimensional spaces
		of four-momentum and spacetime coordinates, or rather with the
		three-dimensional spaces of momenta and positions only.
		Below I argue that there are reasons for choosing
		the second alternative, at least at the beginning. In fact,
		the classical Hamiltonian formalism
		exhibits symmetry between the
		three-dimensional momentum and position 
		coordinates.  
		Time, however, occupies a distinguished position: 
		it is a parameter upon
		which the position and momentum coordinates depend. 
		An even more pronounced difference between time and 
		the position and
		momentum coordinates exists in quantum mechanics, where time is
		still a $c-$number parameter, 
		while space (momentum space) coordinates
		become operators.
		The present-day relativistic field theory,
		which does unite
		special relativity and quantum physics,
		achieves this in a rather formal way
		which may be argued to be unsatisfactory 
		at the conceptual level
		\cite{Wigner1957,Dirac1973,Chew1971,FinkMcCollum1975}.
		Indeed, Finkestein called it a $c-q$ theory,
		a merger of classical and quantum concepts
		\cite{Finkelstein1971}. 
		Although local relativistic field theory is enormously 
		successful, 
		predicting also the existence 
		of experimentally confirmed
		quantum correlations between spatially separated events,
		the entailed issue of an instantaneous quantum reduction 
		of a non-local state vector
		does not seem to be
		in accord with the original {spirit} of relativity
		\cite{PenroseBook}. The latter problem highlights the
		essential difference
		between time and space in spite of these notions being united
		into the concept of the Minkowskian spacetime.
		Furthermore, one has to keep in mind 
		that local Lorentz-equivalent 
		frames are not fully equivalent in Nature:
		the background radiation is isotropic in one such frame only.
		Given this situation I think that it is appropriate 
		to restrict our considerations 
		to the realm of nonrelativistic
		physics hoping that 
		a proper treatment of relativistic effects 
		can be achieved at a later stage.
		
\subsection{Symmetry Transformations}		
		
	The introduction of $\kappa$ permits the combination of separate
	invariants $\bf{p}^2$ and $\bf{x}^2$ into a single form 
	$\bf{p}^2+\bf{x}^2$, in which position and momenta
	 coordinates enter on a completely equal footing,
	and suggests subsequent consideration of all transformations 
	that leave this form invariant.
        These transformations clearly include standard 
	three-dimensional rotations and ordinary reflections.
	As shown in \cite{ZenczyIJTP1990}
	the set of all transformations 
	which leave the form $\bf{p}^2+\bf{x}^2$ invariant, 
	under the restriction that also the Poisson brackets 
	$\{p_i,p_k\}$, $\{x_i,x_k\}$, $\{p_i,x_k\}$  are to be form invariant, 
	forms the group $U(1) \otimes SU(3)$.
	The $U(1)$ factor takes care of reflections ($\bf{p}'=-\bf{p}$,
	$\bf{x}'=-\bf{x}$)
	and the reciprocity transformations of Born ($\bf{p}'=\bf{x}$,
	$\bf{x}'=-\bf{p}$), 
	the former being the squares of the latter;
	while the $SU(3)$ group constitutes the
	extension of the group of proper rotations,
	where the latter are understood as simultaneous rotations of $\bf{p}$ 
	and $\bf{x}$.
	
	In the six-dimensional space of 
	$(p_1,p_2,p_3,p_4,p_5,p_6) \equiv (p_1,p_2,p_3,x_1,x_2,x_3)\equiv 
	{\bf{p}}\oplus{\bf{x}}$
	with the fifteen $SO(6)$ generators $G_{ik}$ ($=-G_{ki}$)
	given by
	\begin{equation}
	(G_{mn})_{ik} = \delta_{mi}\delta_{nk}-\delta_{mk}\delta_{ni},
	\end{equation}
	the eight $SU(3)$ generators are as follows:
	\begin{eqnarray}
	F_1&=-H_3=&G_{15}+G_{24}\\
	F_2&=-J_3=&G_{12}+G_{45}\\
	F_3&=R_1-R_2=&G_{41}-G_{52}\\
	F_4&=-H_2=&G_{34}+G_{16}\\
	F_5&=J_2=&G_{13}+G_{46}\\
	F_6&=H_1=&G_{62}+G_{53}\\
	F_7&=J_1=&G_{32}+G_{65}\\
	F_8&=(R_1+R_2-2R_3)/\sqrt{3}=&(G_{41}+G_{52}-2G_{63})/\sqrt{3}.
	\end{eqnarray}
	These generators satisfy
	 the commutation rules of the $su(3)$ Lie algebra
	\begin{equation}
	\label{SU3commutationrules}
	[F_i,F_k]=2 f_{ikj}F_j
	\end{equation}
	with totally antisymmetric structure constants $f_{ikj}$ equal to $1$
	for
	$ikj=(123)$; $1/2$ for $ikj=(147), (165), (246), (257),$ $(345), (376)$;
	$\sqrt{3}/2$ for $ikj=(458), (678)$; and zero otherwise.
	The $U(1)$ generator is
	\begin{equation}
	R=R_1+R_2+R_3=G_{41}+G_{52}+G_{63}.
	\end{equation}

	The question now emerges 
	in what way this $U(1) \otimes SU(3)$ group should be used to generalize 
	the old notions.
	In contemporary particle physics, 
	when an enlarged symmetry group is introduced, 
	the action of new group generators 
	produces new objects (e.g. fundamental particles)
	from the old ones. Standardly,
	this is being done at the level of a
	smallest-dimensionsal irreducible spinorial 
	representation of the group.
	There is no necessity, however, to employ
	the enlarged symmetry group in exactly this way
	(for a remotely related remark see e.g. \cite{CNYang1973}).

\subsection{Momentum and Mass}

The "${\bf{p}}\oplus {\bf{x}}$" scheme as defined above does not yet really
distinguish between momentum and
position coordinates. 
It is only when
the standard concept of mass is introduced
that a real difference
between $\bf{p}$ and $\bf{x}$ appears.
Namely, we observe that 
for individual objects separated from each other by 
large distances, be it in the macroworld or in the world of elementary particles, 
energy of directly observed free clasical objects (quantum particles) 
is defined by their mass
and {\em momenta}, whether via a relativistic or a nonrelativistic formula.
Thus, the standard concept of mass may be said to be directly associated with
the concept of momentum $\bf{p}$, not position $\bf{x}$. 
In other words,  the six-dimensional vector $(p_1,p_2,...p_6)$ 
is divided into two triplets in such a way 
that one of the triplets ("$\bf{p}$") is associated with mass.

When the "${\bf{p}}\oplus {\bf{x}}$" philosophy 
admitting transformations between $\bf{p}$ and $\bf{x}$
is accepted, one naturally asks
whether this division into momenta coordinates associated with mass,
and position coordinates not associated with this notion is unique, or not.
In fact, 
the "${\bf{p}}\oplus {\bf{x}}$" scheme suggests that 
the above division may not be
unique, and that the
momentum and position coordinates
could be treated on a more equal footing.
I shall now argue how this can be done.
First, however, let us note that
transformations of the $U(1) \otimes SU(3)$ group
were constructed to effect the ${\bf{p}} \leftrightarrow {\bf{x}}$ 
transformations, not 
to act upon mass itself.
Consequently, mass (an invariant of the group of ordinary rotations) 
should also be an invariant of the constructed group.

Within the "${\bf{p}}\oplus {\bf{x}}$" philosophy, 
the (unknown) mechanism generating standard particle masses 
must somehow divide the 6-dimensional object "${\bf{p}}\oplus {\bf{x}}$" 
into a pair $\{\bf{p},\bf{x}\}$ of canonically conjugated
3-dimensional variables
$\bf{p}$ and $\bf{x}$, with one of these ($\bf{p}$) 
directly associated with the
concept of mass.
However, the division of the 6-dimensional object "${\bf{p}}\oplus {\bf{x}}$" 
into two 3-dimensional canonically conjugated objects 
(i.e. into a pair \{(generalized momenta), (generalized positions)\}) 
may proceed in several ways, leading 
not only to the standard form 
\begin{equation}
\label{standardmomentum}
\{(p_1,p_2,p_3),(x_1,x_2,x_3)\},
\end{equation}
but also to 
\begin{equation}
\label{red}
\{(p_1,x_2,-x_3),(x_1,-p_2,p_3)\}
\end{equation}
or to
\begin{equation}
\label{yellow}
 \{(-x_1,p_2,x_3) , (p_1,x_2,-p_3)\}
 \end{equation}
or to 
\begin{equation}
\label{blue}
\{(x_1,-x_2,p_3) , (-p_1,p_2,x_3)\},
\end{equation}
where the signs take into account in particular the requirement
that the Poisson brackets 
of new momenta and positions are to be the same as before.
Clearly, there are other possibilities, including
the cases when all $p_i$ and $x_k$ in the above pairs
are interchanged, such as eg. (for Eq.(\ref{red}))
\begin{equation}
\label{even}
\{(x_1,-p_2,p_3),(-p_1,-x_2,x_3)\}.
\end{equation}
There are two sets of such choices of canonical momenta and positions: 
one with an odd number of standard momentum components 
 in the new canonical momentum as in Eqs (\ref{standardmomentum}-\ref{blue}), 
and one with an even number of standard momentum components, 
as in Eq.(\ref{even}).

All such new pairings of $p_1,p_2,p_3,x_1,x_2,x_3$ into 
the canonically conjugated momenta and positions
may be obtained from the standard pairing $\{\bf{p},\bf{x}\}$
via the action of appropriate elements from the said
$U(1) \otimes SU(3)$ group.
For example, $SU(3)$ rotation by $90^o$ using
the generator $H_1$, followed by an appropriate
ordinary rotation simultaneously in planes
$(x_2,x_3)$ and $(p_2,p_3)$,  leads to 
the first of three pairs given above, i.e. to Eq.(\ref{red}), 
which is also obtained by an
appropriate rotation by
$F_3$.
Similarly, analogous rotation 
using $H_2$ followed by an ordinary rotation or $(F_3 +\sqrt{3}F_8)/2$ 
($H_3$ or $(F_3 -\sqrt{3}F_8)/2$) leads to
Eq.(
\ref{yellow}) (respectively: Eq.(\ref{blue})).

Transformations of the standard pairing $\{\bf{p},\bf{x}\}$ 
using the $SU(3)$ group 
lead therefore to the first set of
additional pairings, i.e. to three-dimensional
generalized momenta in which one component is a component of the 
standard momentum
coordinate, 
while the remaining two components are the components of the standard position
coordinate.
The cases with $\bf{p}$ and $\bf{x}$ interchanged 
(two standard momenta coordinates and one standard position coordinate 
constituting
a generalized
momentum together)
are generated from the former cases by the $U(1)$ generator $R$.

As already mentioned, transformations generated by $R$ lead in particular to 
the reciprocity transformations, i.e. they exchange $\{\bf{p},\bf{x}\}$ 
into $\{\bf{p'},\bf{x'}\}=\{-\bf{x},\bf{p}\}$.
Contrary to the $SU(3)$ transformations generated by $H_i$, the transformations
induced by $R$ are clearly acceptable by the condition of rotational invariance. 
Furthermore, $\bf{p'}^2={\bf{x}}^2$ could also be made acceptable by translational invariance
upon an interpretation of $\bf{x}$ as a difference of position coordinates of
two objects. Despite that, in our macroscopic world
we do not seem to observe objects for which energy and mass  
are connected with the standard position coordinates $\bf{x}$ 
in a way analogous to that in which
they are connected 
with the standard momentum coordinates $\bf{p}$. 
In fact, momentum "$\bf{p}$" was defined as this subtriplet of six variables 
$p_1,p_2,...,x_3$ which associates the concept of mass to energy, 
contrary to the
other subtriplet, "$\bf{x}$".
This suggests that consideration of transformation generated by $R$ 
may lead outside
the realm of objects with masses.

Thus, it is only through the $SU(3)$ factor,
the minimal simple-group extension of 
the group of standard rotations, that one 
 arrives at the proposal for how to generalize the link
between the concepts of energy, mass and ordinary momentum.
While I have no idea how the mass-generating mechanism actually operates,
the $SU(3)$ symmetry present in the "${\bf{p}}\oplus {\bf{x}}$" scheme
suggests a unique way of applying the standard concept of mass 
to 1+3 types of divisions of standard
momenta and position
coordinates into pairs of canonically conjugated triplets.
Accordingly,
there are three additional choices for generalized momenta,
besides the ordinary momentum $\bf{p}$, which may be linked to
the concept of mass.
Each of these three choices clearly violates ordinary rotational
invariance (translational invariance is satisfied if $\bf{x}$ is understood
as denoting the difference of position coordinates).
Thus, the objects (if any)
for which mass is linked with generalized momenta of the type $(p_1,x_2,-x_3)$
certainly cannot belong as individual objects to our macroworld since the latter
is rotationally invariant.
These objects could, however, belong to the macroworld as unseparable
components of composite objects, provided the latter are constructed in such a
way that they satisfy all  invariance
conditions (rotational
etc.) requested.

It is now very tempting to conjecture here that 
the three additional types of objects, linking mass to three choices for
generalized momenta and related to each other by rotations, correspond to
quarks. The unobservability of individual quarks would then be directly related
to this lack of rotational (and possibly translational) invariance.
An argument against the above proposal claiming
that strong interactions are known to be invariant under the
transformations from both the homogenous and inhomogenous Poincar\'e group,
including 3-dimensional translations and rotations, is not a valid one.
Indeed, we do know that any hadron, 
when probed by objects exhibiting all required transformation 
properties (e.g. a photon, a W-boson, etc.) 
does exhibit transformation 
properties compensating those of a probe, 
so that the whole interaction is rotationally and translationally invariant.
However, this {\em always} concerns the interaction with hadron {\em as a whole},
and 
{\em never} with an
{\em individual} quark (in the SM it is the interaction with a {\em colour-singlet}
{\em superposition} of quark currents, which - as far as colour is concerned -
possesses the properties of a hadron, and not those of an individual quark).
If the individual quarks {\it conspire} in ensuring the requested transformation
properties of hadrons, the whole scheme could be a viable one.
Obviously, the conspiration mechanism should ensure that somehow
the quarks of a hadron might be
described by rotationally covariant entities - i.e. ordinary spinors. 
The difference with the standard approach would then be that the
objects of definite mass are not the objects with standard 
spinorial transformation properties.
This type of property is well known in the Standard Model where 
quark states in which weak currents are diagonal and quark states in which
quark mass matrix is diagonal are not the same, but related by a unitary
transformation.
It may be that the only essential difference of the present
 proposal with respect to the
SM is the relaxation of the standard way of treating quark masses.
However, further changes might also be needed.

Symmetry and simplicity of the scheme
make me find it hard to believe that Nature has not utilized 
the above possibility. 
Rather, the problem seems to me to be
 how to put the above ideas into an appropriate mathematical form, and how to
 develop the latter.
 In the next sections  a simple construction will be proposed,
 which satisfies
 certain of the requirements discussed above. 

This construction is of a $c-q$ type in the classification of Finkelstein, and
thus is most probably a great oversimplification, in particular when the
complexity of hadron physics is recalled. 
Yet, it exhibits a series of interesting qualitative properties and
admits a simple interpretation. Although it does have some shortcomings
and is presumably a toy model only, I think it is worth presenting.

\section{Generalizing Dirac Hamiltonian}	

The content of the previous section
suggests that the basic inputs of present theories, which are based
on the standard link between the concept of mass and momentum,
should be appropriately generalized. In particular this concerns the Dirac equation.
Indeed, no $SU(3)$ "${\bf{p}}\oplus {\bf{x}}$" symmetry 
is present in the Dirac equation:
when the latter is written down in momentum representation, it is 
completely oblivious to space and time.
It exhibits connection with space(-time)
   only through the quantum/wave route
   with the help of the Planck constant of dimension $GeV\cdot cm$.
   When restricted to momentum-space representation 
   the Dirac Hamiltonian 
  does not exhibit any quantum features, its properties being of a purely
   geometric classical nature.

Below we shall start with the Dirac Hamiltonian in momentum space 
and then proceed to act on it
with the $SU(3)$
transformations discussed above. Now, one may argue that there is an
inconsistency here: the
Dirac equation is relativistic, while the proposed phase-space approach is 
nonrelativistic. For our purposes, however,
what is important in the Dirac Hamiltonian is 1) its algebra, and 2)
the fact that it leads to antiparticles.
Now, as far as the algebra is concerned, the 
Dirac's trick of doubling the size of matrices (from $2 \times 2$ to 
$4 \times 4$), although originally needed to linearize the form
${\bf{p}}^2+m^2$,
is also needed to represent reflections (a
nonrelativistic concept, needed in the ${\bf{p}}^2+{\bf{x}}^2$ scheme as well),
 since these
cannot be described by Pauli matrices alone.
Thus, the matrices
\begin{eqnarray}
\alpha_k & = & \left[ 
\begin{array}{cc} 0 & \sigma _k \\
                 \sigma_k & 0 \end{array}
\right] =
\sigma _k \otimes \sigma _1 \\
\beta & = &  \left[
\begin{array}{cc} 1 & 0 \\
                 0 & -1 \end{array}
\right]=
\sigma_0 \otimes \sigma_3, 
\end{eqnarray}
with $\sigma _k$ being Pauli matrices,
and $\beta $ needed to represent the reflection
through
$\alpha_k \to \beta \alpha_k \beta = -\alpha _k$,
should be appropriate in our nonrelativistic case as well.

The term $\alpha _k p_k$ is rotation- and reflection- invariant and is clearly
appropriate for the linearization of both the relativistic and nonrelativistic 
expressions for Hamiltonians, both containing $\bf{p}^2$.
It is only when $\beta $, multiplied by the mass parameter $m$, is added to 
$\alpha _k p_k$ and the whole expression is identified with a Hamiltonian, i.e.
\begin{equation}
H_D = \alpha _k p_k + \beta m,
\end{equation}
that one recovers a relativistically covariant expression.
We will use the above Hamiltonian not only 
because of its simplicity when compared to the
analogous Hamiltonian obtained when one linearizes the 
Schr\"odinger equation \cite{LevyLeblond1971},
but mainly because 
it leads {\em in a standard way} to antiparticles and
the well-known procedure of charge conjugation which we shall need. 
What is important here is the form of the relativistic relation 
to be linearized,
in which energy, mass and momentum enter in squares,
unlike in the nonrelativistic case for which 
both energy and mass enter in a linear
fashion: 
\begin{equation}
\label{schrod}
2 m E= {\bf{p}}^2.
\end{equation} 
Should one insist on a manifestly nonrelativistic treatment,
application of $SU(3)$ transformations to the linearized Schr\"o\-dinger equation  
proceeds in a way fairly analogous to the Dirac case discussed below.
 It is only when charge
conjugation (hence antiparticles) is considered 
that a difference in the treatment of mass is needed
(see Section \ref{chargeconj}).
Alternatively, we may argue that we do not really 
need the full relativistic-invariance of the Dirac Hamiltonian:
we may restrict ourselves to
very large ratios of mass to momentum, when the Dirac Hamiltonian may be
considered as approximating the strictly
nonrelativistic case with the important addition that it  involves
negative energy solutions to be interpreted later as
antiparticles.

In order to proceed with $SU(3)$ transformations on the Dirac Hamiltonian 
we have to introduce $\bf{x}$ 
in addition to $\bf{p}$ (and $m$), and linearize the ${\bf{p}}^2+{\bf{x}}^2$ 
$(+m^2)$ 
form, still treating $\bf{p}$ and $\bf{x}$ as commuting variables. In order to do this 
we have to enlarge Dirac matrices by doubling their size and introducing
\begin{eqnarray}
A_k=&\alpha_k \otimes \sigma_0 
&= \sigma _k \otimes \sigma _1 \otimes \sigma _0 \\
B  =&\beta \otimes \sigma _0 
&= \sigma_0 \otimes \sigma _3 \otimes \sigma _0 \\
\label{Bk}
B_k=&&=\sigma_0 \otimes \sigma_2 \otimes \sigma _k
\end{eqnarray}
with matrices $B_k$ associated with $x_k$.
The above matrices satisfy the conditions:
\begin{eqnarray}
A_k A_l + A_l A_K & = & 2 \delta _{kl}\\
\label{AkBl}
A_k B_l + B_l A_K & = & 0 \\
B_k B_l + B_l B_K & = & 2 \delta _{kl}\\
A_k B +B A_k      & = & 0\\
B_k B +B B_k      & = & 0 \\
B B &=& 1
\end{eqnarray}
Note that because of the requirement that all
the $p_i x_k$ mixed terms are to vanish, and that $\bf{x}$ (and hence
$B_k$) changes sign under reflection, $B_k$ has to contain 
$\sigma_2$ as a second factor in the tensor product in Eq.(\ref{Bk}).

Applying now the  $SU(3)$ transformation induced by $H_1$ (followed by a
rotation in (2,3)-planes), i.e.
\begin{eqnarray}
(p_1,p_2,p_3;x_1,x_2,x_3)& \to & (p_1,x_2,-x_3;x_1,-p_2,p_3)\\
(A_1,A_2,A_3;B_1,B_2,B_3)& \to & (A_1,B_2,-B_3;B_1,-A_2,A_3)
\end{eqnarray}

we arrive at the following counterpart of the Dirac Hamiltonian:
\begin{equation}
\label{redquark}
H_R=A_1p_1+B_2x_2+B_3x_3+Bm.
\end{equation}
This Hamiltonian is not invariant under standard rotations, 
and - if $x_2,x_3$ are not
position differences - it is not translationally invariant
either. Restoration of rotational (and/or translational) symmetry would require
addition of appropriate terms. We shall discuss this in due time.
Now, in line with the three possible ways of choosing generalized momenta
(cf. Eqs.(\ref{red},\ref{yellow},\ref{blue})) 
two further Hamiltonians may be constructed
apart from the Hamiltonian of Eq.(\ref{redquark}), namely:
\begin{equation}
\label{yellowquark}
H_Y=B_1x_1+A_2p_2+B_3x_3+Bm
\end{equation}
and
\begin{equation}
\label{bluequark}
H_B=B_1x_1+B_2x_2+A_3p_3+Bm
\end{equation}
The subscripts distinguish betwen these Hamiltonians, 
and were thought of as abbreviations for "Red", "Yellow",
and "Blue", proposed in anticipation that the above
Hamiltonians will perhaps turn out appropriate for the description of quarks.

Note that under the above proposal the
$SU(3)$ group in question, which offers a generalization of the rotation group, 
does provide a set of transformations leading 
from the old objects to the new ones. 
However, this occurs at the level of a pair of
vectors $\bf{p}$ and $\bf{x}$ (or axial vectors, because $SU(3)$ is not concerned
with reflections),  not
 at the level of the spinorial
representation of the rotation group \cite{CNYang1973}.

\subsection{Restoring Rotational Invariance}
Consider now an ordinary rotation of the reference frame
around the third axis by an arbitrary angle $\phi$.
Matrices $\bf{A}$ and $\bf{B}$ transform under this rotation 
in the same way as
vectors $\bf{p}$ and $\bf{x}$. Below we denote the transformed matrices and vectors
with a prime sign. 
Hamiltonian $H_B$ remains then form-invariant:
\begin{equation}
H_B \equiv H_B(x_1,x_2,p_3) = H'_B(x'_1,x'_2,p'_3),
\end{equation}
where the prime sign in $H'_B$ refers to the transformed matrices $A'_k$, $B'_k$,
whereas Hamiltonians $H_R$ and $H_Y$ acquire the following look:
\begin{eqnarray}
\label{HRtransformed}
H_R(p_1,x_2,x_3)&=&c^2 H'_R(p'_1,x'_2,x'_3)+s^2H'_Y(x'_1,p'_2,x'_3)\\
&&+sc (B'_2x'_1+B'_1x'_2-A'_2p'_1-A'_1p'_2)\nonumber \\
\label{HYtransformed}
H_Y(x_1,p_2,x_3)&=&c^2 H'_Y(x'_1,p'_2,x'_3)+s^2H'_R(p'_1,x'_2,x'_3)\\
&&-sc (B'_2x'_1+B'_1x'_2-A'_2p'_1-A'_1p'_2),\nonumber 
\end{eqnarray}
with $s = \sin \phi$, $c = \cos \phi$.
The sum of the Hamiltonians $H_R+H_Y$ is clearly form-invariant:
\begin{equation}
H_R(p_1,x_2,x_3)+H_Y(x_1,p_2,x_3)=H'_R(p'_1,x'_2,x'_3)+H'_Y(x'_1,p'_2,x'_3),
\end{equation}
which is true if  $p_1,p_2$ ($x_1,x_2$) 
constitute two components of a {\em single} vector $\bf{p}$
(respectively $\bf{x}$), as is implicit on the r.h.s. of Eqs. 
(\ref{HRtransformed},\ref{HYtransformed}) .
It is now obvious that the Hamiltonian:
\begin{equation}
\label{fullrotinvH}
H=H_R(p_1,x_2,x_3)+H_Y(x_1,p_2,x_3)+H_B(x_1,x_2,p_3)
\end{equation}
is form-invariant under arbitrary three-dimensional rotations.
Consequently, with the Hamiltonian as above, it does not matter
whether we perform the division of the six-dimensional 
object $\{\bf{p},\bf{x}\}$
into the three pairs of triplets of Eqs(\ref{red},\ref{yellow},\ref{blue}) 
of a given frame of reference
\begin{eqnarray}
\label{redorig}
(R)&\phantom{xxxxxx}&\{(p_1,x_2,-x_3),(x_1,-p_2,p_3)\}\\
\label{yelloworig}
(Y)&& \{(-x_1,p_2,x_3) , (p_1,x_2,-p_3)\}\\
\label{blueorig}
(B)&&\{(x_1,-x_2,p_3) , (-p_1,p_2,x_3)\},
\end{eqnarray}
or do this
in the rotated frame for $\{{\bf{p}}',{\bf{x}}'\}$ 
according to:
\begin{eqnarray}
\label{redprime}
(R')&\phantom{xxxxxx}&\{(p'_1,x'_2,-x'_3),(x'_1,-p'_2,p'_3)\}\\
\label{yellowprime}
(Y')&& \{(-x'_1,p'_2,x'_3) , (p'_1,x'_2,-p'_3)\}\\
\label{blueprime}
(B')&&\{(x'_1,-x'_2,p'_3) , (-p'_1,p'_2,x'_3)\}.
\end{eqnarray}
In the above considerations $x_1$ ($x_2$, $x_3$) occurs in two
Hamiltonians: $H_Y$ and $H_B$ ($H_R$ and $H_B$, $H_R$ and $H_Y$).
In principle one could introduce here two different position-type vectors
$\bf{x}$ and $\bf{y}$ so that each of the three Hamiltonians $H_R,H_Y,H_B$ would
depend on three independent components of generalized momenta:
$H_R(p_1,x_2,y_3)$, $H_Y(y_1,p_2,x_3)$, and $H_B(x_1,y_2,p_3)$.
The analog of Eq.(\ref{fullrotinvH}) would be still rotationally invariant.
Although this extension might be relevant, it is not needed
for the presentation of our main idea. 
Consequently, below we shall have one position-type 
vector only ($\bf{x}$).
If the requirement of translational invariance is imposed, 
$\bf{x}$ has to be understood as a difference of position coordinates.
We shall come back to this issue in Section \ref{qantiq}.

\subsection{Charge Conjugation}
\label{chargeconj}
As mentioned earlier, the Dirac Hamiltonian 
illustrates our ideas probably in the simplest way, in particular
permitting also the standard introduction of the operation of
charge conjugation. 
In order to discuss the latter along the standard lines we need to replace the
up-to-now classical variables of $\bf{p}$ and $\bf{x}$ with operators.

First, we recall how charge conjugation operation is
effected when electromagnetic interaction is added to the Dirac Hamiltonian:
\begin{equation}
\label{HDwithA}
H_D=A_k (p_k-e {\cal{A}}_k) +B m + e {\cal{A}}_0,
\end{equation}
where ${\cal{A}}_{\mu}$ denotes the electromagnetic field.  

In order to transform $H_D$ to a form in which its antiparticle content
is explicit 
(i.e. in which the positive energy solution corresponds to antiparticles) 
we have to change the relative sign between $p_k$ and $e
{\cal{A}}_{k}$.
This is standardly obtained via complex conjugation applied to both
c-numbers and operators with the properties: 
$i \to -i$, ${\bf{x}} \to {\bf{x}}$, $t \to t$, $H \to -H$, 
${\bf{p}} \to -{\bf{p}}$, $A_k \to A^*_k$
($A^*_{1,3}=A_{1,3}$, $A^*_2=-A_2$), $B \to B^*=B$, 
${\cal{A}}_{\mu}\to {\cal{A}}^*_{\mu}={\cal{A}}_{\mu}$, and $m \to m$, 
$e \to e$
[in order to describe antiparticles 
when linearizing the Schr\"odinger equation  one would
have to
use $m \to -m$ (instead of $m \to m$) when $E \to -E$ (c.f. Eq.(\ref{schrod})].
Application of these rules to Eq.(\ref{HDwithA}) leads to
\begin{equation}
\label{HDcomplex}
H'=A^*_k (p_k+e {\cal{A}}_k) -B m - e {\cal{A}}_0.
\end{equation}
When the following matrix is introduced ($\tau ^2=1$):
\begin{equation}
C = - i \sigma_2 \otimes \sigma_2\otimes \tau = - C^{-1},
\end{equation}
which satisfies
\begin{eqnarray}
C B C^{-1}&=&-B\\
C A^*_k C^{-1}&=&A_k, 
\end{eqnarray}
the above Hamiltonian may be transformed to the standard form with an opposite
sign of charge:
\begin{equation}
H' \to C H' C^{-1}= A_k (p_k+e {\cal{A}}_k) +B m - e {\cal{A}}_0,
\end{equation}
as appropriate for the explicit description of antiparticles.

Now, when complex conjugation (defined as above) 
is applied to $H_R$ (with $B_k \to B^*_k$
and $B^*_{1,3}=-B_{1,3}$, $B^*_2=B_2$), 
one obtains the analogue of Eq. (\ref{HDcomplex}) (we skipped the
electromagnetic field)
\begin{equation}
H'_R=A_1 p_1-B_2 x_2 +B_3 x_3 - B m.
\end{equation}
For $H_Y$ and $H_B$ one similarly obtains:
\begin{eqnarray}
H'_Y&=&B_1 x_1+A_2 p_2 +B_3 x_3 - B m,\\
H'_B&=&B_1 x_1-B_2 x_2 +A_3 p_3 - B m.
\end{eqnarray}

Choosing now 
\begin{equation}
\tau = \sigma_2
\end{equation}
which leads to:
\begin{eqnarray}
C B^*_k C^{-1}&=&B_k\\
&\rm{or}&\nonumber \\
C B_{1,3} C^{-1}&=& - B_{1,3}\\
C B_2 C^{-1}&=&B_2,
\end{eqnarray}
we find that $H'_R$, $H'_Y$, and $H'_B$
are transformed to
\begin{eqnarray}
H_{{\bar{R}}}  &=& A_1 p_1 -B_2 x_2 - B_3 x_3 + B m,\\
H_{{\bar{Y}}} &=& - B_1 x_1 + A_2 p_2 - B_3 x_3 + B m,\\
H_{{\bar{B}}} &=& - B_1 x_1 - B_2 x_2 + A_3 p_3 + B m,
\end{eqnarray}
(Choices $\tau = \sigma_0, \sigma_1, \sigma_3$ do not lead to
a rotationally invariant form simultaneously for all three
$B x$ terms in $H_{\bar{R}}, H_{\bar{Y}}, H_{\bar{B}}$.)
Note that neither an ordinary rotation nor reflection ($A_k \to B A_k B
= -A_k$; $B_k \to B B_k B= -B_k$)
can bring $H_{\bar{R}}$ into $H_R$ 
(i.e. $A_1p_1 \to A_1p_1$ and $-B_3x_3 \to +B_3x_3$ ?) etc.
Thus, even without considering the charge explicitly, 
$H_{\bar{R}}$ represents an object different from that described by $H_R$.
In summary, when we are given a Hamiltonian for a particle, 
we form a Hamiltonian for its antiparticle by replacing $e$ and $\bf{x}$ 
with $-e$ and $-\bf{x}$ while keeping the rest of the particle Hamiltonian
unchanged.

\subsection{Quark-Antiquark Systems}
\label{qantiq}
We are now in a position to consider a Hamiltonian for a  
quark-antiquark system. In order to construct
this Hamiltonian one has to add the Hamiltonians of system components. 
Here one encounters a problem whether the matrices $A_i$, $B_k$ for quarks 
and those for antiquarks are distinct or identical.
The simplest possibility is that they are identical.
Here, one may point out a parallel to the current-field identity of hadronic
physics, according to which the electromagnetic vector current
$j_{\mu}=\bar{q}\gamma_{\mu}q$, involving matrices $\gamma _i \leftrightarrow
BA_i$ is identical to the vector meson field, built out of $q\bar{q}$ pair. 
Consequently,
we propose to add the Hamiltonians of red quarks and antiquarks as follows:
\begin{equation}
H_R+H_{\bar{R}}=A_1(p_1+\bar{p}_1) +B_2(x_2-\bar{x}_2)+B_3(x_3-\bar{x}_3)+2~B~m,
\end{equation}
where bars over momenta or positions identify these as {\em physical}
antiquark variables. 
Given the complexity of hadronic physics even in the meson spectrum only,
this proposal is most probably an oversimplification, but it exhibits several
interesting features.
For $x_k$ ($\bar{x}_k$)
we may now admit the position coordinates themselves and not
their differences only.
(If an overall position coordinate $X_1$, canonically conjugate
to $P_1=p_1+{\bar{p}}_1$ from $H_R+H_{\bar{R}}$, 
were included in the definitions of $x_1$ and
${\bar{x}_1}$ in $H_Y$, $H_{\bar{Y}}$, $H_B$, and $H_{\bar{B}}$ through
$x_1 \to x_1-X_1$, ${\bar{x}}_1 \to {\bar{x}_1}-X_1$, 
the dependence on $X_1$ would cancel in $H_Y+H_{\bar{Y}}$
and $H_B+H_{\bar{B}}$ anyway.)
The total rotationally and translationally invariant
Hamiltonian $H_{q\bar{q}}$ of a quark-antiquark system
is therefore:
\begin{equation}
\label{rottransinv}
H_{q\bar{q}}=H_R+H_{\bar{R}}+H_Y+H_{\bar{Y}}+H_B+H_{\bar{B}}=
{\bf{A}} \cdot {\bf{P}}+2~{\bf{B}}\cdot{\Delta \bf{x}}+6 ~B~ m,
\end{equation}
where $\bf{P}$ denotes the total momentum of a system, and $\Delta {\bf{x}}$ -
quark-antiquark "position difference".  
Thus, one has to add contributions from all three colours and from both quarks
and antiquarks to obtain a rotationally and translationally invariant
Hamiltonian.
Hamiltonian (\ref{rottransinv}) 
provides a particular
mathematical realization of the idea of conspiration mechanism
which, as I believe, ensures that the individual 
rotationally and translationally non-invariant quarks
combine to form fully acceptable states.

If we divide the total Hamiltonian into quark and antiquark contributions $H_q$
and $H_{\bar{q}}$, 
we observe that while either of them lacks translational invariance,
it is rotationally invariant, e.g.
\begin{equation}
\label{Hquarks}
H_q={\bf{A}}\cdot {\bf{p}} + {\bf{B}} \cdot 2 {\bf{x}} + 3 ~B~ m.
\end{equation}
With the first and third factors in the direct product 
of $ 2 \times 2$ matrices
in the explicit expressions for ${\bf{A}},{\bf{B}},B$ 
transforming under rotations in the same way,    
this Hamiltonian should be appropriate for a
description of a system of total spin 0 or 1.
However, its coupling to a photon goes solely through the first factor,
i.e. is that of a spin 1/2 object, as needed for a quark.

Let us also point out that
with 
\begin{equation}
\label{gamma5}
\gamma_5\equiv -i A_1 A_2 A_3 = 
\sigma_0 \otimes \sigma _1 \otimes \sigma_0,
\end{equation}
anticommuting with both $B$ and
$\bf{B}$, both the $B m$ and ${\bf{B}}
\cdot {\bf{x}}$ terms are not chirally-invariant, a property
 exhibited by mass terms in standard formulations.
  It is perhaps also worth noting
 here 
  that the analogues of definition (\ref{gamma5}), appropriate
for individual "coloured" quarks, and defined in their respective
subspaces through:
\begin{eqnarray}
\gamma _{R5} =& -iA_1B_2B_3= &\sigma _1 \otimes \sigma_1 \otimes
\sigma_1\\
\gamma _{Y5} =& -iA_2B_3B_1= &\sigma _2 \otimes \sigma_1 \otimes
\sigma_2\\
\gamma _{B5} =& -iA_3B_1B_2= &\sigma _3 \otimes \sigma_1 \otimes
\sigma_3
\end{eqnarray}
anticommute with the quark mass terms $B m$ as well, parallelling a
similar property of $\gamma _5$ 
(however, the full meaning of these properties is not clear).

Translational invariance requires that we add the quark and
antiquark contributions $H_{q\bar{q}}=H_q+H_{\bar{q}}$ as proposed above. 
This procedure 
maintains  0 and 1 as possible values for the total spin of
the composite system,
and satisfies the condition
of the additivity of quark and antiquark charges.

Then, taking the square of $H_{q\bar{q}}$ \'a la Dirac, we obtain
\begin{equation}
\label{mesonmass}
E^2={\bf{p}}^2+4~{\Delta \bf{x}}^2 + (6m)^2,
\end{equation}
with the $4~{\Delta \bf{x}}^2 + (6m)^2$ playing the r\^{o}le of meson mass squared.

There are several interesting qualitative features appearing
in a rudimentary form in the above construction, with some of the important ones
being:
\begin{itemize}
\item Additivity of quark charges. As usual, the charges of individual quarks
have to be adjusted so as to lead
to correct meson charges.
\item Additivity of quark masses. First quark models often had such a property
built in by hand. Later it was viewed as an approximation.
\item Appearance of a "string" between points describing the "locations" of
a quark and an antiquark. This was an argument in favour of introducing 
$\kappa $ of dimension $GeV/cm $. 
An analogous constant (when Planck constant is added) appears in 
the original dual string model of mesons
\cite{Veneziano,Rebbi}, which
introduces  a constant of the dimension of $cm^2$ 
in its definition of string action. 
This dual string model exhibits various features in qualitative agreement with
the observed properties of hadronic amplitudes.
\item Objects exhibiting well-defined properties of one type do not have
well-defined properties of another type.
\end{itemize}

For the description of baryons one needs an extension of the
algebra of $A_k$, $B_k$ matrices (still with SU(3) symmetry properties). 
Indeed, in the construction discussed before
there are only two $2 \times 2$ matrix factors transforming under rotation in 
the standard way, while three such factors are needed to admit the description
of the spin of three baryon quarks. The corresponding matrices should be
multiplied by the  available translationally-invariant three-vectors.
Thus,
in addition to the triplet of
momentum coordinates one needs
two triplets of position coordinate differences.
However, from six canonical position coordinates ($\bf{x}$, $\bf{y}$) 
in general present 
in the three Hamiltonians
$H_R,H_Y,H_B$ one can form only one triplet of position differences
${\bf{x}}-{\bf{y}}$.
Consequently, it is not clear to me how 
the baryon-describing Hamiltonian should be constructed.
Perhaps one should use the same triplet of position differences 
in both spinorial
subspaces.
This would suggest that in baryons 
one space degree of freedom is actually frozen. 
Such a possibility was discussed at length in baryon phenomenology
\cite{linear,KoniukIsgur}, and
is a viable option in light of the experimental information.
Or maybe the overall position coordinate $\bf{X}$ 
canonically conjugated to the total momentum $\bf{P}$
should be somehow introduced 
(a kind of analogue of the "string junction" in the string model of baryons
\cite{junction}). A deeper insight on
how to proceed with baryons would be very helpful.

\section{Final Remarks}
The approach of the preceeding Section proposes a particular mathematical
realization of the arguments of Section \ref{generalization}. It is formulated
at the $c-q$ level and for this reason it is presumably still a
 toy model. However, it exhibits several very interesting qualitative
properties, which should follow in a natural way from any more involved
approach including the possibly underlying quantum pregeometry.

The considered scheme exhibits some (though somewhat
superficial)
similarities with the rishon model
of leptons and quarks \cite{Harari}, in which leptons of a given generation
(e.g. $e^+$, ${\nu_{e}}$)
are constructed from 
rishons $T$ (of charge +1/3) and $V$ (of charge 0) as $TTT$
and $VVV$, while coloured quarks of the same generation,  
i.e. $u_r,u_y,u_b$, and $\bar{d}_r,\bar{d}_y,\bar{d}_b$
are the (ordered) composites 
$VTT$, $TVT$, $TTV$ and $TVV$, $VTV$, $VVT$ respectively.
The rishon model suggests that our scheme should be doubled to
incorporate weak isospin.
In line with the general idea that the attributes of elementary particles 
should be somehow linked 
with the properties of the classical arena used for the
description of Nature, one should
try to propose a correspondence betwen this doubling and phase-space properties. 
Most probably it should be related to
a relative relation between the momentum and position coordinates, 
as our scheme, contrary to the standard approach,
admits considering their independent transformations.
However, at this stage we prefer not to speculate
on this subject any further. Still, it should be mentioned that the 
rishon model was incorporated in a much more elaborated scheme involving
the standard $SU(3)_C \times U(1)_{EM}$ gauge invariance \cite{HarariSeiberg}.

I hope that the
 proposed approach provides an interesting suggestion concerning 
 the unobservability of quarks and the concept of quark mass. 
 Obviously, the ideas put forward herein
 may constitute a mirage only.
 Should these ideas turn out to be relevant, one could look
 further, guided by the conjecture that
 the problem of mass is intimately connected or even
 identical to the problem of space (phase-space) quantization.

\vfill

\vfill

\end{document}